\documentclass[a4paper,11pt]{article}

\usepackage[dvips]{hyperref}
\hypersetup{
colorlinks=true,
linkcolor=magenta,
citecolor=blue,
urlcolor=black,
bookmarksnumbered=true,
bookmarkstype=toc}
\usepackage[margin=1in]{geometry}
\usepackage{amsmath,amsbsy}
\usepackage[adobe-utopia]{mathdesign}
\usepackage[T1]{fontenc}
\usepackage{cite}
\usepackage{fancyhdr}
\makeatletter

\@addtoreset{equation}{section}
\makeatother
\usepackage{titletoc}
\titlecontents{section}[1.5em]{\bfseries\mathversion{bold}}{\contentslabel{1.5em}}{\hspace*{-1.5em}}{\titlerule*[0pc]{}\contentspage}[\addvspace{1ex}]
\usepackage{titlesec}
\titleformat{\section}[block]{\filright\bfseries\mathversion{bold}}{\thesection.}{0.5em}{}[\titlerule]
\titleformat{\subsection}[block]{\filright\bfseries\mathversion{bold}}{\thesubsection.}{0.5em}{}

\title{
\bf\large A Simple Derivation of Finite-Temperature CFT Correlators\\ from the BTZ Black Hole
}
\author{
Satoshi Ohya${}^{1,2}$\\[2ex]
\textit{\small ${}^{1}$Department of Physics, Faculty of Nuclear Sciences and Physical Engineering}\\
\textit{\small Czech Technical University in Prague}\\
\textit{\small Pohrani\v{c}n\'{i} 1288/1, 40501 D\v{e}\v{c}\'{i}n, Czech Republic}\\
\textit{\small ${}^{2}$Doppler Institute for Mathematical Physics and Applied Mathematics}\\
\textit{\small Czech Technical University in Prague}\\
\textit{\small B\v{r}ehov\'{a} 7, 11519 Prague, Czech Republic}\\[2ex]
\texttt{\small Email:\,\href{mailto:ohyasato@fjfi.cvut.cz}{ohyasato@fjfi.cvut.cz}}
}
\date{\small (Dated: \today)}

\begin{document}
\maketitle
\thispagestyle{fancy}
\renewcommand{\headrulewidth}{0pt}
\rhead{DI13-063}

\begin{abstract}
We present a simple Lie-algebraic approach to momentum-space two-point functions of two-dimensional conformal field theory at finite temperature dual to the BTZ black hole.
Making use of the real-time prescription of AdS/CFT correspondence and ladder equations of the Lie algebra $\mathfrak{so}(2,2) \cong \mathfrak{sl}(2,\mathbb{R})_{L} \oplus \mathfrak{sl}(2,\mathbb{R})_{R}$, we show that the finite-temperature two-point functions in momentum space satisfy linear recurrence relations with respect to the left and right momenta.
These recurrence relations are exactly solvable and completely determine the momentum-dependence of retarded and advanced two-point functions of finite-temperature conformal field theory.
\end{abstract}

\tableofcontents

\newpage
\section{Introduction and summary} \label{sec:1}
Conformal symmetry is powerful enough to constrain the possible forms of correlation functions in quantum field theory.
It has been long appreciated that, for scalar (quasi-)primary operators, for example, $SO(2,d)$ conformal symmetry completely fixes the possible forms of two- and three-point functions up to an overall normalization factor in any spacetime dimension $d \geq 1$.
This symmetry constraint works well in position space, however, its direct implication to momentum-space correlators are less obvious before perfuming Fourier transform.
Since momentum-space correlators are directly related to physical observables (e.g. imaginary part of retarded two-point function in momentum space gives spectral density of many body systems), it is important to understand how directly conformal symmetry constrains the possible forms of momentum-space correlators.
From practical computational viewpoint, it is also important to develop efficient methods to compute momentum-space correlators directly through symmetry considerations, because Fourier transforms of position-space correlators are hard in general.

In this short paper we continue our investigation \cite{Ohya:2013xva} and present a novel Lie-algebraic approach to momentum-space two-point functions of conformal field theory at finite temperature by using the AdS/CFT correspondence.
The AdS/CFT correspondence relates strongly-coupled conformal field theory to classical gravity in one-higher spatial dimension.
According to the correspondence, finite-temperature conformal field theory is dual to an asymptotically AdS spacetime that contains black holes.
In this paper we focus on two-dimensional conformal field theory (CFT$_{2}$) at finite temperature dual to the three-dimensional anti-de Sitter (AdS$_{3}$) black hole (i.e. Ba\~{n}ados-Teitelboim-Zanelli (BTZ) black hole \cite{Banados:1992wn,Banados:1992gq}) and give a simple derivation of retarded and advanced two-point functions of scalar operators of dual CFT$_{2}$ by just using the real-time prescription of AdS/CFT correspondence \`{a} la Iqbal and Liu \cite{Iqbal:2008by,Iqbal:2009fd} and the ladder equations of the Lie-algebra $\mathfrak{so}(2,2) \cong \mathfrak{sl}(2,\mathbb{R})_{L} \oplus \mathfrak{sl}(2,\mathbb{R})_{R}$ of the isometry group $SO(2,2) \cong (SL(2,\mathbb{R})_{L} \times SL(2,\mathbb{R})_{R})/\mathbb{Z}_{2}$ of AdS$_{3}$.
In contrast to the conventional approaches to momentum-space CFT correlators (such as Fourier-transform of position-space correlators or original real-time AdS/CFT prescription \cite{Son:2002sd,Iqbal:2008by,Iqbal:2009fd} that requires to solve bulk field equations explicitly), our Lie-algebraic method is quite simple and clarifies the role of conformal symmetry in momentum-space correlators in a direct way: For finite-temperature two-point functions in momentum space, conformal symmetry manifests itself in a form of recurrence relations, which are exactly solvable and, up to an overall normalization factor, completely determine the momentum dependence of two-point functions.

The rest of the paper is organized as follows.
In section \ref{sec:2} we briefly review the AdS$_{3}$ black hole based on the quotient construction \cite{Banados:1992gq,Steif:1995zm}: The AdS$_{3}$ black hole is a locally AdS$_{3}$ spacetime and given by a quotient space of AdS$_{3}$ with an identification of points under the action of a discrete subgroup $\mathbb{Z}$ of the isometry group $SO(2,2)$ of AdS$_{3}$.
Though not so widely appreciated, the AdS$_{3}$ black hole is a quotient space of AdS$_{3}$ with a particular coordinate patch in which both the time- and angle-translation generators generate the one-parameter subgroup $SO(1,1) \subset SO(2,2)$.\footnote{This is true for non-extremal black hole with positive mass. The time- and angle-translation generators generate other one-parameter subgroups for the zero-mass limit of black hole (or black hole vacuum), the extremal black hole and the negative mass black hole (or black hole with naked singularity). For example, in the case of the black hole vacuum, the time- and angle-translation generators generate the subgroup $E(1) \times E(1) \subset SO(2,2) \cong (SL(2,\mathbb{R})_{L} \times SL(2,\mathbb{R})_{R})/\mathbb{Z}_{2}$ prior to making the $\mathbb{Z}$-identification. (Note that $SL(2,\mathbb{R})$ contains three distinct one-parameter subgroups: the rotation group $SO(2)$, the Lorentz group $SO(1,1)$ and the Euclidean group $E(1)$.) For detailed discussions of the quotient construction, we refer to the original paper \cite{Banados:1992gq} (see also \cite{Steif:1995zm}).}
In section \ref{sec:3} we introduce a coordinate realization of the Lie algebra $\mathfrak{so}(2,2) \cong \mathfrak{sl}(2,\mathbb{R})_{L} \oplus \mathfrak{sl}(2,\mathbb{R})_{R}$ realized in scalar field theory on the AdS$_{3}$ black hole background.
We then demonstrate in section \ref{sec:4} a simple Lie-algebraic method to compute the retarded and advanced CFT$_{2}$ two-point functions by just using the ladder equations of the Lie algebra $\mathfrak{so}(2,2) \cong \mathfrak{sl}(2,\mathbb{R})_{L} \oplus \mathfrak{sl}(2,\mathbb{R})_{R}$ in the basis in which $SO(1,1) \times SO(1,1) \subset SO(2,2)$ generators become diagonal.
We will see that our method correctly reproduces the known results \cite{Birmingham:2001pj,Iqbal:2009fd,Balasubramanian:2010sc}.

\section{\texorpdfstring{AdS$_{3}$}{AdS3} black hole: Locally \texorpdfstring{AdS$_{3}$}{AdS3} spacetime in the \texorpdfstring{$SO(1,1) \times SO(1,1)$}{SO(1,1) x SO(1,1)} diagonal basis} \label{sec:2}
Let us start with the following non-rotating BTZ black hole described by the metric
\begin{align}
ds_{\text{AdS$_{3}$ BH}}^{2}
&= 	- \left(\frac{\rho^{2}}{R^{2}} - 1\right)d\tau^{2}
	+ \frac{d\rho^{2}}{\rho^{2}/R^{2} - 1}
	+ \rho^{2}d\theta^{2}, \label{eq:2.1}
\end{align}
where $\tau \in (-\infty, +\infty)$, $\rho \in (0,\infty)$, $\theta \in [0, 2\pi)$, and $R > 0$ is the AdS$_{3}$ radius.
In this paper we simply call \eqref{eq:2.1} the AdS$_{3}$ black hole and focus on the region outside the horizon $\rho > R$.
For the following discussions it is convenient to introduce a new spatial coordinate $x$ as follows:
\begin{align}
\rho
&= 	R\coth(x/R), \label{eq:2.2}
\end{align}
where $x$ ranges from $0$ to $\infty$.
Notice that the black hole horizon $\rho = R$ corresponds to $x = \infty$, while the AdS$_{3}$ boundary $\rho = \infty$ corresponds to $x = 0$.
A straightforward calculation shows that in the coordinate system $(\tau, x, \theta)$ the black hole metric \eqref{eq:2.1} takes the following form:
\begin{align}
ds_{\text{AdS$_{3}$ BH}}^{2}
&= 	\frac{-d\tau^{2} + dx^{2} + R^{2}\cosh^{2}(x/R)d\theta^{2}}{\sinh^{2}(x/R)}. \label{eq:2.3}
\end{align}
For the sake of notational brevity, we will hereafter work in the units in which $R=1$.

Several comments are in order:
\begin{enumerate}
\item \textbf{BTZ black hole.}
The above AdS$_{3}$ black hole \eqref{eq:2.1} is locally isometric to the rotating BTZ black hole \cite{Banados:1992wn,Banados:1992gq} and obtained by suitable change of spacetime coordinates.
Indeed, it is easy to show that the BTZ black hole metric
\begin{align}
ds_{\text{BTZ}}^{2}
&= 	- \frac{(r^{2} - r_{+}^{2})(r^{2} - r_{-}^{2})}{r^{2}}dt^{2}
	+ \frac{r^{2}dr^{2}}{(r^{2} - r_{+}^{2})(r^{2} - r_{-}^{2})}
	+ r^{2}\left(d\phi - \frac{r_{+}r_{-}}{r^{2}}dt\right)^{2}, \label{eq:2.4}
\end{align}
where $r_{+}$ and $r_{-}$ are outer and inner horizons, respectively, is reduced to the AdS$_{3}$ black hole \eqref{eq:2.1} by the following coordinate change \cite{KeskiVakkuri:1998nw}:
\begin{align}
\rho
= 	\sqrt{\frac{r^{2} - r_{-}^{2}}{r_{+}^{2} - r_{-}^{2}}}, \quad
\tau
= 	r_{+}t - r_{-}\phi, \quad
\theta
= 	r_{+}\phi - r_{-}t. \label{eq:2.5}
\end{align}
Notice that the light-cone coordinates satisfy the relations $\tau \pm \theta = (r_{+} \mp r_{-})(t \pm \phi)$.

\item \textbf{\mathversion{bold}Local coordinate patch of AdS$_{3}$.}
The AdS$_{3}$ black hole is a locally AdS$_{3}$ spacetime and obtained from the AdS$_{3}$ spacetime with a suitable periodic identification \cite{Banados:1992gq,Steif:1995zm}.
To see this, let us first note that the AdS$_{3}$ spacetime can be embedded into the four-dimensional ambient space $\mathbb{R}^{2,2}$ and defined as the following hypersurface with constant negative curvature $-1 (= -1/R^{2})$:
\begin{align}
\text{AdS}_{3}
&= 	\left\{
	(X^{-1}, X^{0}, X^{1}, X^{2}) \in \mathbb{R}^{2,2}
	\mid
	- (X^{-1})^{2} - (X^{0})^{2} + (X^{1})^{2} + (X^{2})^{2} = -1
	\right\}. \label{eq:2.6}
\end{align}
The AdS$_{3}$ black hole \eqref{eq:2.3} is given by the following local coordinate patch of the hypersurface:
\begin{align}
(X^{-1}, X^{0}, X^{1}, X^{2})
&= 	\left(
	\coth x\cosh\theta,
	\frac{\sinh\tau}{\sinh x},
	\coth x\sinh\theta,
	\frac{\cosh\tau}{\sinh x}
	\right), \label{eq:2.7}
\end{align}
with the periodic identification $\theta \sim \theta + 2n\pi$ ($n \in \mathbb{Z}$).
In fact, it is straightforward to show that the induced metric $ds_{\text{AdS}_{3}}^{2} = \left.- (dX^{-1})^{2} - (dX^{0})^{2} + (dX^{1})^{2} + (dX^{2})^{2}\right|_{(X^{-1}, X^{0}, X^{1}, X^{2}) \in \text{AdS}_{3}}$ on the hypersurface takes the following form:
\begin{align}
ds_{\text{AdS}_{3}}^{2}
&= 	\frac{-d\tau^{2} + dx^{2} + \cosh^{2}x d\theta^{2}}{\sinh^{2}x}. \label{eq:2.8}
\end{align}
It should be emphasized that the periodic identification $\theta \sim \theta + 2n\pi$ makes the metric \eqref{eq:2.8} black hole.
As mentioned in \cite{Banados:1992gq}, without such identification the metric \eqref{eq:2.8} just describes a portion of AdS$_{3}$ and the horizon is just that of an accelerated observer.
(Roughly speaking, \eqref{eq:2.7} is the AdS$_{3}$ counterpart of Rindler coordinate patch of Minkowski spacetime.)

\item \textbf{\mathversion{bold}$SO(1,1) \times SO(1,1)$ global symmetry.}
As is well-known, the AdS$_{3}$ spacetime \eqref{eq:2.6} has an alternative equivalent description as the $SL(2, \mathbb{R})$ group manifold defined as follows:
\begin{align}
\text{AdS}_{3}
&= 	\left\{
	\left.
	X
	= 	\begin{pmatrix}
		X^{-1} + X^{2} 	& X^{1} - X^{0} \\
		X^{1} + X^{0} 	& X^{-1} - X^{2}
		\end{pmatrix}
	~\right\vert~
	\det X = 1
	\right\}. \label{eq:2.9}
\end{align}
With this definition it is obvious that the AdS$_{3}$ spacetime \eqref{eq:2.9} is invariant under left- and right-multiplications of $SL(2,\mathbb{R})$ matrices, $X \mapsto X^{\prime} = g_{L}Xg_{R}$, where $g_{L} \in SL(2,\mathbb{R})_{L}$ and $g_{R} \in SL(2,\mathbb{R})_{R}$ with the $\mathbb{Z}_{2}$-identification $(g_{L}, g_{R}) \sim (-g_{L}, -g_{R})$.
(Notice that $(g_{L}, g_{R})$ and $(-g_{L}, -g_{R})$ give the same $X^{\prime}$.)
In the local coordinate patch \eqref{eq:2.7} the $2 \times 2$ matrix $X = \left(\begin{smallmatrix} X^{-1} + X^{2} & X^{1} - X^{0} \\ X^{1} + X^{0} & X^{-1} - X^{2}\end{smallmatrix}\right)$ takes the following form:
\begin{align}
X
= 	\begin{pmatrix}
	\displaystyle \coth x\cosh\theta + \frac{\cosh\tau}{\sinh x} &
	\displaystyle \coth x\sinh\theta - \frac{\sinh\tau}{\sinh x} \\[1em]
	\displaystyle \coth x\sinh\theta + \frac{\sinh\tau}{\sinh x} &
	\displaystyle \coth x\cosh\theta - \frac{\cosh\tau}{\sinh x}
	\end{pmatrix}. \label{eq:2.10}
\end{align}
Now it is easy to see that the time-translation $\tau \mapsto \tau^{\prime} = \tau + \epsilon$ is induced by the noncompact $SO(1,1) \subset SO(2,2)$ group action $X \mapsto X^{\prime} = g_{L}Xg_{R}$ given by the matrices
\begin{align}
g_{L}
&= 	g_{R}^{-1}
= 	\begin{pmatrix}
	\cosh\frac{\epsilon}{2}	& \sinh\frac{\epsilon}{2} \\[.3em]
	\sinh\frac{\epsilon}{2} 	& \cosh\frac{\epsilon}{2}
	\end{pmatrix}
	\in SO(1,1). \label{eq:2.11}
\end{align}
Likewise, the spatial-translation $\theta \mapsto \theta^{\prime} = \theta + \epsilon$ is induced by another $SO(1,1)  \subset SO(2,2)$ group action $X \mapsto X^{\prime} = g_{L}Xg_{R}$ given by the matrices
\begin{align}
g_{L}
&= 	g_{R}
= 	\begin{pmatrix}
	\cosh\frac{\epsilon}{2}	& \sinh\frac{\epsilon}{2} \\[.3em]
	\sinh\frac{\epsilon}{2} 	& \cosh\frac{\epsilon}{2}
	\end{pmatrix}
	\in SO(1,1). \label{eq:2.12}
\end{align}
Hence, prior to making the periodic identification $\theta \sim \theta + 2n\pi$, the time- and spatial-translation generators $i\partial_{\tau}$ and $-i\partial_{\theta}$ must be given by two distinct $SO(1,1)$ generators of the Lie group $SO(2,2) \cong (SL(2,\mathbb{R})_{L} \times SL(2,\mathbb{R})_{R})/\mathbb{Z}_{2}$.
After the periodic identification $\theta \sim \theta + 2n\pi$, on the other hand, $g_{L}$ and $g_{R}$ in Eq.~\eqref{eq:2.12} should be regarded as an element of the coset $SO(1,1)/\mathbb{Z}$,\footnote{Notice that the parameter space of $SO(1,1)$ (more precisely, $SO_{+}(1,1)$, i.e. the connected component to the identity element) is the whole line $\mathbb{R}$. Hence the parameter space of $SO(1,1)/\mathbb{Z}$ is $\mathbb{R}/\mathbb{Z}$, which is isomorphic to a circle $S^{1}$.} where the $\mathbb{Z}$-identification is defined by $\epsilon \sim \epsilon + 2n\pi$ ($n \in \mathbb{Z}$).
Hence, the AdS$_{3}$ black hole is given by the quotient space AdS$_{3}$/$\mathbb{Z}$, where the identification subgroup $\mathbb{Z} = \{g^{n} \mid n=0,\pm1,\pm2,\cdots\} \subset SO(2,2)$ is generated by the matrix $g = g_{L} = g_{R} = \left(\begin{smallmatrix}\cosh\pi & \sinh\pi \\ \sinh\pi & \cosh\pi\end{smallmatrix}\right)$.

It should be emphasized here that the fact that the time-translation generator generates the noncompact Lorentz group $SO(1,1)$ is a manifestation of thermodynamic aspects of black hole: If we work in Euclidean signature, the noncompact Lorentz group $SO(1,1)$ becomes the compact rotation group $SO(2) \cong S^{1}$ such that the frequencies conjugate to the imaginary time are quantized and hence give rise to the Matsubara frequencies.

\item \textbf{Two-point function.}
As we have seen, the AdS$_{3}$ black hole \eqref{eq:2.1} is a locally AdS$_{3}$ spacetime but its global structure is quite different from AdS$_{3}$.
This global difference of course leads to a big difference between the structure of two-point functions of CFT$_{2}$ living on the boundary of AdS$_{3}$ black hole and those living on the boundary of AdS$_{3}$ \cite{KeskiVakkuri:1998nw}.
To see this, let $G_{\text{AdS$_{3}$ BH}}(\tau, \theta)$ be a scalar two-point function of CFT$_{2}$ dual to the AdS$_{3}$ black hole and $G_{\text{AdS$_{3}$}}(\tau, \theta)$ be a scalar two-point function of CFT$_{2}$ living on the AdS$_{3}$ boundary without periodic identification.
Then, once we get $G_{\text{AdS$_{3}$}}(\tau, \theta)$, the scalar two-point function of CFT$_{2}$ dual to the AdS$_{3}$ black hole is given by the coset construction (or the method of images \cite{KeskiVakkuri:1998nw}):
\begin{align}
G_{\text{AdS$_{3}$ BH}}(\tau, \theta)
&= 	\sum_{n\in\mathbb{Z}}\rho(n)G_{\text{AdS$_{3}$}}(\tau, \theta + 2n\pi), \label{eq:2.13}
\end{align}
where $\rho: \mathbb{Z} \to U(1)$ is a scalar (i.e. one-dimensional) unitary representation of the identification subgroup $\mathbb{Z}$ and given by $\rho(n) = \mathrm{e}^{in\alpha}$.
Here $\alpha$ is a real parameter and its value depends on the model.
For example, for scalar operator $\mathcal{O}(\tau, \theta)$ that satisfies the periodic boundary condition $\mathcal{O}(\tau, \theta + 2\pi) = \mathcal{O}(\tau, \theta)$, $\alpha$ is zero (i.e. $\rho$ is the trivial representation).
(Basically, $\alpha$ is a boundary condition parameter for $\mathcal{O}(\tau,\theta)$ with respect to the angle $\theta$.)
We emphasize that, regardless of the value of $\alpha$, thus constructed two-point function \eqref{eq:2.13} indeed satisfies the periodic boundary condition $G_{\text{AdS$_{3}$ BH}}(\tau, \theta + 2\pi) = G_{\text{AdS$_{3}$ BH}}(\tau, \theta)$.

For simplicity throughout this paper we will focus on $G_{\text{AdS$_{3}$}}$ (i.e. the zero-winding sector of $G_{\text{AdS$_{3}$ BH}}$), because $G_{\text{AdS$_{3}$ BH}}$ can be constructed from the knowledge of $G_{\text{AdS$_{3}$}}$.
Hence in what fallows we do not need to worry about the subtleties of periodic identification and global difference between the AdS$_{3}$ black hole and the AdS$_{3}$ spacetime.\footnote{Actually, the momentum-space two-point functions computed in Refs.~\cite{Birmingham:2001pj,Iqbal:2009fd,Balasubramanian:2010sc} are nothing but the momentum-space representation of $G_{\text{AdS}_{3}}$ rather than $G_{\text{AdS$_{3}$ BH}}$ (or $G_{\text{BTZ}}$).}
\end{enumerate}

Let us next consider a massive scalar field $\phi$ of mass $m$ on the background spacetime \eqref{eq:2.8} (without periodic identification) that satisfies the Klein-Gordon equation $(\Box_{\text{AdS}_{3}} - m^{2})\phi = 0$, where the d'Alembertian is given by
$\Box_{\text{AdS}_{3}}
= 	\sinh^{2}x
	\left[
	- \partial_{\tau}^{2} + \partial_{x}^{2} - \frac{1}{\sinh x\cosh x}\partial_{x}
	- \frac{-\partial_{\theta}^{2}}{\cosh^{2}x}
	\right]$.
In order to get CFT two-point functions via real-time AdS/CFT prescription, we need to find a solution to the Klein-Gordon equation whose $\tau$- and $\theta$-dependences are given by the plane waves, $\phi(\tau, x, \theta) = \phi_{\omega, k}(x)\mathrm{e}^{-i\omega\tau+ik\theta}$; that is, we need to know a \textit{simultaneous eigenfunction} of the d'Alembertian $\Box_{\text{AdS}_{3}}$, the time-translation generator $i\partial_{\tau}$ and the spatial-translation generator $-i\partial_{\theta}$.
For such a simultaneous eigenfunction the Klein-Gordon equation reduces to the following differential equation:\footnote{Redefining the field as $\phi \mapsto \Tilde{\phi} = (\coth x)^{-1/2}\phi$, one sees that the differential equation \eqref{eq:2.13} reduces to the Schr\"{o}dinger equation with hyperbolic P\"{o}schl-Teller potential
\begin{align}
\left(
- \partial_{x}^{2}
+ \frac{(\Delta-1)^{2} - 1/4}{\sinh^{2}x} + \frac{k^{2} + 1/4}{\cosh^{2}x}
\right)
\Tilde{\phi}_{\omega, k}
&= 	\omega^{2}\Tilde{\phi}_{\omega, k}. \nonumber
\end{align}}
\begin{align}
\left(
- \partial_{x}^{2} + \frac{1}{\sinh x\cosh x}\partial_{x}
+ \frac{\Delta(\Delta - 2)}{\sinh^{2}x} + \frac{k^{2}}{\cosh^{2}x}
\right)
\phi_{\omega, k}
&= 	\omega^{2}\phi_{\omega, k}, \label{eq:2.14}
\end{align}
where $\Delta = 1 + \sqrt{m^{2} + 1}$ is one of the solutions to the quadratic equation $\Delta(\Delta - 2) = m^{2}$.
Notice that near the AdS$_{3}$ boundary $x=0$ the differential operator in the left-hand side of \eqref{eq:2.13} behaves as $-\partial_{x}^{2} + \frac{1}{x}\partial_{x} + \frac{\Delta(\Delta-2)}{x^{2}} + O(1)$.
Hence the general solution has the following asymptotic near-boundary behavior:
\begin{align}
\phi(\tau, x, \theta)
&\sim
	A_{\Delta}(\omega, k)x^{\Delta}\mathrm{e}^{-i\omega\tau + ik\theta}
	+ B_{\Delta}(\omega, k)x^{2-\Delta}\mathrm{e}^{-i\omega\tau + ik\theta}
	\quad\text{as}\quad
	x \to 0, \label{eq:2.15}
\end{align}
where $A_{\Delta}(\omega, k)$ and $B_{\Delta}(\omega, k)$ are integration constants which may depend on $\Delta$, $\omega$ and $k$.
The real-time prescription of AdS/CFT correspondence tells us that the retarded and advanced two-point functions are given by the ratio \cite{Iqbal:2008by,Iqbal:2009fd}
\begin{align}
G_{\Delta}^{R/A}(\omega, k)
&= 	(2\Delta - 2)\frac{A_{\Delta}(\omega, k)}{B_{\Delta}(\omega, k)}, \label{eq:2.16}
\end{align}
where the retarded two-point function $G_{\Delta}^{R}$ is obtained by the solution that satisfies the in-falling boundary conditions at the horizon, whereas the advanced two-point function $G_{\Delta}^{A}$ is obtained by the solution that satisfies the out-going boundary conditions at the horizon \cite{Son:2002sd}.

The goal of this paper is to compute the ratio \eqref{eq:2.15} in a Lie-algebraic fashion without solving the Klein-Gordon equation explicitly.

\section{Lie algebra \texorpdfstring{$\mathfrak{sl}(2,\mathbb{R})_{L} \oplus \mathfrak{sl}(2,\mathbb{R})_{R}$}{sl(2,R)L + sl(2,R)R} in the \texorpdfstring{$SO(1,1) \times SO(1,1)$}{SO(1,1) x SO(1,1)} diagonal basis} \label{sec:3}
In order to get the momentum-space two-point functions,  we need to find a simultaneous eigenfunction of the d'Alembertian $\Box_{\text{AdS}_{3}}$, the time-translation generator $i\partial_{\tau}$ and the spatial-translation generator $-i\partial_{\theta}$.
As we will see below, the d'Alembertian is given by the quadratic Casimir of the Lie algebra $\mathfrak{so}(2,2) \cong \mathfrak{sl}(2,\mathbb{R})_{L} \oplus \mathfrak{sl}(2,\mathbb{R})_{R}$.
On the other hand, as we have seen in the previous section, the time- and spatial-translations are induced by two distinct noncompact $SO(1,1)$ group actions such that $i\partial_{\tau}$ and $-i\partial_{\theta}$ must be given by $SO(1,1)$ generators of the Lie algebra $\mathfrak{so}(2,2) \cong \mathfrak{sl}(2,\mathbb{R})_{L} \oplus \mathfrak{sl}(2,\mathbb{R})_{R}$.
Hence we need to work in the basis in which the noncompact $SO(1,1)$ generators become diagonal.
We note that unitary representations of the Lie algebra $\mathfrak{sl}(2,\mathbb{R})$ in the noncompact $SO(1,1)$ basis have been studied in the mathematical literature \cite{Kariyan:1968,Lindblad:1970}, and known to be a bit complicated.
In this paper we will not touch upon these mathematical subtleties and not discuss which of the unitary representations are realized in the scalar field theory on the background \eqref{eq:2.8} (without periodic identification).\footnote{One way to avoid these subtleties is to Wick-rotate \textit{both} the time $\tau$ and angle $\theta$. In such Euclidean-like signature, the noncompact $SO(1,1) \times SO(1,1)$ symmetry becomes the compact $SO(2) \times SO(2)$ symmetry such that we can use standard unitary representations of the Lie-algebra $\mathfrak{so}(2,2) \cong \mathfrak{sl}(2,\mathbb{R})_{L} \oplus \mathfrak{sl}(2,\mathbb{R})_{R}$ in the $SO(2) \times SO(2)$ diagonal basis. In this approach, computations of momentum-space two-point functions are essentially reduced to those presented in Ref.~\cite{Ohya:2013xva}.}
Instead, we will present a rather heuristic argument that reproduces the known results by just using the ladder equations of the Lie algebra $\mathfrak{so}(2,2)$ in the $SO(1,1) \times SO(1,1)$ diagonal basis.

To begin with, let us first recall the Lie algebra $\mathfrak{so}(2,2) \cong \mathfrak{sl}(2,\mathbb{R})_{L} \oplus \mathfrak{sl}(2,\mathbb{R})_{R}$, which is spanned by six self-adjoint generators $\{A_{0}, A_{1}, A_{2}, B_{0}, B_{1}, B_{2}\}$ that satisfy the following commutation relations:
\begin{subequations}
\begin{alignat}{5}
&[A_{0}, A_{1}] = iA_{2},&\quad
&[A_{1}, A_{2}] = -iA_{0},&\quad
&[A_{2}, A_{0}] = iA_{1},& \label{eq:3.1a}\\
&[B_{0}, B_{1}] = iB_{2},&\quad
&[B_{1}, B_{2}] = -iB_{0},&\quad
&[B_{2}, B_{0}] = iB_{1},& \label{eq:3.1b}
\end{alignat}
\end{subequations}
with other commutators vanishing, $[A_{a}, B_{b}] = 0$ ($a,b = 0,1,2$).
We note that $A_{0}$ and $B_{0}$ are compact $SO(2)$ generators, whereas $A_{1}$, $A_{2}$, $B_{1}$, $B_{2}$ are noncompact $SO(1,1)$ generators.
Note also that the standard classification of unitary representations of the Lie algebra $\mathfrak{so}(2,2)$ is based on the Cartan-Weyl basis $\{A_{0}, A_{1} \pm iA_{2}, B_{0}, B_{1} \pm iB_{2}\}$, where $A_{1} \pm iA_{2}$ and $B_{1} \pm iB_{2}$ play the role of ladder operators that raise and lower the eigenvalues of $A_{0}$ and $B_{0}$ by $\pm 1$.
For the following discussions, however, it is convenient to introduce the \textit{hermitian} linear combinations $A_{\pm} = A_{2} \mp A_{0}$ and $B_{\pm} = B_{2} \mp B_{0}$, which also play the role of ``ladder'' operators; see next section.
In the basis $\{A_{1}, A_{\pm}, B_{1}, B_{\pm}\}$ the commutation relations \eqref{eq:3.1a} and \eqref{eq:3.1b} are cast into the following forms:
\begin{subequations}
\begin{alignat}{3}
&[A_{1}, A_{\pm}] = \pm iA_{\pm},&\quad
&[A_{+}, A_{-}] = 2iA_{1},& \label{eq:3.2a}\\
&[B_{1}, B_{\pm}] = \pm iB_{\pm},&\quad
&[B_{+}, B_{-}] = 2iB_{1}.& \label{eq:3.2b}
\end{alignat}
\end{subequations}
In the problem of scalar field theory on the background \eqref{eq:2.8} (without periodic identification), these symmetry generators are turned out to be given by the following first-order differential operators:
\begin{subequations}
\begin{align}
A_{1}
&= 	\frac{i}{2}(\partial_{\tau} + \partial_{\theta}), \label{eq:3.3a}\\
A_{\pm}
&= 	-\frac{i}{2}\mathrm{e}^{\pm(\tau+\theta)}
	\left[
	\sinh x\partial_{x} \pm \left(\cosh x\partial_{\tau} + \frac{1}{\cosh x}\partial_{\theta}\right)
	\right], \label{eq:3.3b}\\
B_{1}
&= 	\frac{i}{2}(\partial_{\tau} - \partial_{\theta}), \label{eq:3.3c}\\
B_{\pm}
&= 	+\frac{i}{2}\mathrm{e}^{\pm(\tau-\theta)}
	\left[
	\sinh x\partial_{x} \pm \left(\cosh x\partial_{\tau} - \frac{1}{\cosh x}\partial_{\theta}\right)
	\right], \label{eq:3.3d}
\end{align}
\end{subequations}
which indeed satisfy the commutation relations \eqref{eq:3.2a} and \eqref{eq:3.2b}.
The quadratic Casimir of the Lie algebra $\mathfrak{so}(2,2) \cong \mathfrak{sl}(2,\mathbb{R})_{L} \oplus \mathfrak{sl}(2,\mathbb{R})_{R}$ yields the d'Alembertian on the AdS$_{3}$ black hole
\begin{align}
C_{2}(\mathfrak{so}(2,2))
&= 	2C_{2}(\mathfrak{sl}(2,\mathbb{R})_{L}) + 2C_{2}(\mathfrak{sl}(2,\mathbb{R})_{R}) \nonumber\\
&= 	\sinh^{2}x
	\left(
	- \partial_{\tau}^{2} + \partial_{x}^{2}
	- \frac{1}{\sinh x\cosh x}\partial_{x} - \frac{-\partial_{\theta}^{2}}{\cosh^{2}x}
	\right), \label{eq:3.4}
\end{align}
where the quadratic Casimir of each $\mathfrak{sl}(2,\mathbb{R})$ is given by $C_{2}(\mathfrak{sl}(2,\mathbb{R})_{L}) = (A_{0})^{2} - (A_{1})^{2} - (A_{2})^{2} = -A_{1}(A_{1} \pm i) - A_{\mp}A_{\pm}$ and $C_{2}(\mathfrak{sl}(2,\mathbb{R})_{R}) = (B_{0})^{2} - (B_{1})^{2} - (B_{2})^{2} = -B_{1}(B_{1} \pm i) - B_{\mp}B_{\pm}$.
A straightforward calculation shows that $C_{2}(\mathfrak{sl}(2,\mathbb{R})_{L})$ and $C_{2}(\mathfrak{sl}(2,\mathbb{R})_{R})$ coincide and are given by
\begin{align}
C_{2}(\mathfrak{sl}(2,\mathbb{R})_{L})
= 	C_{2}(\mathfrak{sl}(2,\mathbb{R})_{R})
= 	\frac{1}{4}C_{2}(\mathfrak{so}(2,2)). \label{eq:3.5}
\end{align}

\paragraph{Asymptotic near-boundary algebra.}
We are interested in the asymptotic near-boundary behavior of the solution to the Klein-Gordon equation \eqref{eq:2.14}.
To analyze this, let us introduce the boundary symmetry generators defined as the limit $x\to0$ of \eqref{eq:3.3a}--\eqref{eq:3.3d}:
\begin{subequations}
\begin{align}
A_{1}^{0}
&:= 	\lim_{x\to0}A_{1}
= 	\frac{i}{2}(\partial_{\tau} + \partial_{\theta}), \label{eq:3.6a}\\
A_{\pm}^{0}
&:= 	\lim_{x\to0}A_{\pm}
= 	-\frac{i}{2}\mathrm{e}^{\pm(\tau+\theta)}
	\left[
	x\partial_{x} \pm \left(\partial_{\tau} + \partial_{\theta}\right)
	\right], \label{eq:3.6b}\\
B_{1}^{0}
&:= 	\lim_{x\to0}B_{1}
= 	\frac{i}{2}(\partial_{\tau} - \partial_{\theta}), \label{eq:3.6c}\\
B_{\pm}^{0}
&:= 	\lim_{x\to0}B_{\pm}
= 	+\frac{i}{2}\mathrm{e}^{\pm(\tau-\theta)}
	\left[
	x\partial_{x} \pm \left(\partial_{\tau} - \partial_{\theta}\right)
	\right], \label{eq:3.6d}
\end{align}
\end{subequations}
which still satisfy the commutation relations of the Lie algebra $\mathfrak{so}(2,2) \cong \mathfrak{sl}(2,\mathbb{R})_{L} \oplus \mathfrak{sl}(2,\mathbb{R})_{R}$
\begin{subequations}
\begin{alignat}{3}
&[A_{1}^{0}, A_{\pm}^{0}] = \pm iA_{\pm}^{0},&\quad
&[A_{+}^{0}, A_{-}^{0}] = 2iA_{1}^{0},& \label{eq:3.7a}\\
&[B_{1}^{0}, B_{\pm}^{0}] = \pm iB_{\pm}^{0},&\quad
&[B_{+}^{0}, B_{-}^{0}] = 2iB_{1}^{0}.& \label{eq:3.7b}
\end{alignat}
\end{subequations}
The quadratic Casimir of this asymptotic near-boundary algebra, which we denote by $\mathfrak{so}(2,2)^{0} \cong \mathfrak{sl}(2,\mathbb{R})_{L}^{0} \oplus \mathfrak{sl}(2,\mathbb{R})_{R}^{0}$, takes the following simple form:
\begin{align}
C_{2}(\mathfrak{so}(2,2)^{0})
= 	4C_{2}(\mathfrak{sl}(2,\mathbb{R})_{L}^{0})
= 	4C_{2}(\mathfrak{sl}(2,\mathbb{R})_{R}^{0})
= 	x^{2}\partial_{x}^{2} - x\partial_{x}
=: 	C_{2}^{0}. \label{eq:3.8}
\end{align}
As we have repeatedly emphasized, we are interested in simultaneous eigenstates of the d'Alembertian $\Box_{\text{AdS}_{3}} \stackrel{x\to0}{\to} C_{2}^{0}$, the time-translation generator $i\partial_{\tau} = B_{1}^{0} + A_{1}^{0}$ and the spatial-translation generator $-i\partial_{\theta} = B_{1}^{0} - A_{1}^{0}$.
Let $|\Delta, k_{L}, k_{R}\rangle^{0}$ be a simultaneous eigenstate of $C_{2}^{0}$, $A_{1}^{0}$ and $B_{1}^{0}$ that satisfies the following eigenvalue equations:
\begin{subequations}
\begin{align}
C_{2}^{0}|\Delta, k_{L}, k_{R}\rangle^{0}
&= 	\Delta(\Delta - 2)|\Delta, k_{L}, k_{R}\rangle^{0}, \label{eq:3.9a}\\
A_{1}^{0}|\Delta, k_{L}, k_{R}\rangle^{0}
&= 	k_{L}|\Delta, k_{L}, k_{R}\rangle^{0}, \label{eq:3.9b}\\
B_{1}^{0}|\Delta, k_{L}, k_{R}\rangle^{0}
&= 	k_{R}|\Delta, k_{L}, k_{R}\rangle^{0}. \label{eq:3.9c}
\end{align}
\end{subequations}
In the coordinate realization these eigenvalue equations become the following differential equations:
\begin{subequations}
\begin{align}
\left(- \partial_{x}^{2} + \frac{1}{x}\partial_{x}
+ \frac{\Delta(\Delta-2)}{x^{2}}\right)\phi_{\Delta, k_{L}, k_{R}}^{0}
&= 	0, \label{eq:3.10a}\\
\left(i\partial_{x_{L}} - k_{L}\right)\phi_{\Delta, k_{L}, k_{R}}^{0}
&= 	0, \label{eq:3.10b}\\
\left(i\partial_{x_{R}} - k_{R}\right)\phi_{\Delta, k_{L}, k_{R}}^{0}
&= 	0, \label{eq:3.10c}
\end{align}
\end{subequations}
where $x_{L}$ and $x_{R}$ are light-cone coordinates given by $x_{L} = \tau + \theta$ and $x_{R} = \tau - \theta$, and $(k_{L}, k_{R})$ and $(\omega, k)$ are related by $k_{L} = (\omega - k)/2$ and $k_{R} = (\omega + k)/2$.
These differential equations are easily solved with the result
\begin{align}
\phi_{\Delta, k_{L}, k_{R}}^{0}(\tau, x, \theta)
&= 	A_{\Delta}(k_{L}, k_{R})x^{\Delta}
	\mathrm{e}^{-ik_{L}x_{L}}\mathrm{e}^{-ik_{R}x_{R}}
	+ B_{\Delta}(k_{L}, k_{R})x^{2-\Delta}
	\mathrm{e}^{-ik_{L}x_{L}}\mathrm{e}^{-ik_{R}x_{R}}, \label{eq:3.11}
\end{align}
which precisely coincides with the asymptotic near-boundary behavior of the solution \eqref{eq:2.14}.

\section{Recurrence relations for finite-temperature \texorpdfstring{CFT$_{2}$}{CFT2} two-point functions} \label{sec:4}
As mentioned in the previous section, $A_{\pm}$ and $B_{\pm}$ (and also $A_{\pm}^{0}$ and $B_{\pm}^{0}$) play the role of ``ladder'' operators.
To see this, let us consider states $A_{\pm}^{0}|\Delta, k_{L}, k_{R}\rangle^{0}$ and $B_{\pm}^{0}|\Delta, k_{L}, k_{R}\rangle^{0}$.
The commutation relations $[A_{1}^{0}, A_{\pm}^{0}] = \pm iA_{\pm}^{0}$ and $[B_{1}^{0}, B_{\pm}^{0}] = \pm iB_{\pm}^{0}$ give $A_{1}^{0}A_{\pm}^{0}|\Delta, k_{L}, k_{R}\rangle^{0} = (k_{L} \pm i)A_{\pm}^{0}|\Delta, k_{L}, k_{R}\rangle^{0}$ and $B_{1}^{0}B_{\pm}^{0}|\Delta, k_{L}, k_{R}\rangle^{0} = (k_{R} \pm i)B_{\pm}^{0}|\Delta, k_{R}, k_{L}\rangle^{0}$, which imply that $A_{\pm}^{0}$ and $B_{\pm}^{0}$ raise and lower the eigenvalues $k_{L}$ and $k_{R}$ by $\pm i$:\footnote{One may wonder why the eigenvalues of the self-adjoint operators $A_{1}^{0}$ and $B_{1}^{0}$ take the complex values $k_{L} \pm i$ and $k_{R} \pm i$. The reason is that, even if the state $|\Delta, k_{L}, k_{R}\rangle^{0}$ lies inside the domain in which the operators $A_{1}^{0}$ and $B_{1}^{0}$ become self-adjoint, the states $A_{\pm}^{0}|\Delta, k_{L}, k_{R}\rangle^{0}$ and $B_{\pm}^{0}|\Delta, k_{L}, k_{R}\rangle^{0}$ turn out to lie outside the self-adjoint domain of $A_{1}^{0}$ and $B_{1}^{0}$.
(For rigorous mathematical discussions we refer to the literature \cite{Kariyan:1968,Lindblad:1970}.)
As we will see below, however, a naive use of the ``ladder'' equations \eqref{eq:4.1a} and \eqref{eq:4.1b} correctly yields the retarded and advanced two-point functions.}
\begin{subequations}
\begin{align}
A_{\pm}^{0}|\Delta, k_{L}, k_{R}\rangle^{0}
&\propto 	|\Delta, k_{L} \pm i, k_{R}\rangle^{0}, \label{eq:4.1a}\\
B_{\pm}^{0}|\Delta, k_{L}, k_{R}\rangle^{0}
&\propto 	|\Delta, k_{L}, k_{R} \pm i\rangle^{0}. \label{eq:4.1b}
\end{align}
\end{subequations}
In the coordinate realization \eqref{eq:3.6a} and \eqref{eq:3.6c} with the solution \eqref{eq:3.11}, the left-hand sides become
\begin{subequations}
\begin{align}
A_{\pm}^{0}\phi_{\Delta, k_{L}, k_{R}}^{0}
&= 	i\left(-\frac{\Delta}{2} \pm ik_{L}\right)A_{\Delta}(k_{L}, k_{R})
	x^{\Delta}\mathrm{e}^{-i(k_{L} \pm i)x_{L}}\mathrm{e}^{-ik_{R}x_{R}} \nonumber\\
& 	\quad
	+ i\left(\frac{\Delta}{2} - 1 \pm ik_{L}\right)B_{\Delta}(k_{L}, k_{R})
	x^{2-\Delta}\mathrm{e}^{-i(k_{L} \pm i)x_{L}}\mathrm{e}^{-ik_{R}x_{R}}, \label{eq:4.2a}\\
B_{\pm}^{0}\phi_{\Delta, k_{L}, k_{R}}^{0}
&= 	-i\left(-\frac{\Delta}{2} \pm ik_{R}\right)A_{\Delta}(k_{L}, k_{R})
	x^{\Delta}\mathrm{e}^{-ik_{L}x_{L}}\mathrm{e}^{-i(k_{R} \pm i)x_{R}} \nonumber\\
& 	\quad
	- i\left(\frac{\Delta}{2} - 1 \pm ik_{R}\right)B_{\Delta}(k_{L}, k_{R})
	x^{2-\Delta}\mathrm{e}^{-ik_{L}x_{L}}\mathrm{e}^{-i(k_{R} \pm i)x_{R}}, \label{eq:4.2b}
\end{align}
\end{subequations}
which should be proportional to $\phi_{\Delta, k_{L} \pm i, k_{R}}^{0}$ and $\phi_{\Delta, k_{L}, k_{R} \pm i}^{0}$, respectively.
In other words, the integration constants should satisfy the recurrence relations $(-\frac{\Delta}{2} \pm ik_{L})A_{\Delta}(k_{L}, k_{R}) \propto A_{\Delta}(k_{L} \pm i, k_{R})$ and $(\frac{\Delta}{2} -1 \pm ik_{L})B_{\Delta}(k_{L}, k_{R}) \propto B_{\Delta}(k_{L} \pm i, k_{R})$, and similar expressions for $k_{R}$.
Hence the two-point function $G_{\Delta}(k_{L}, k_{R})$, which is given by the ratio $G_{\Delta}(k_{L}, k_{R}) = (2\Delta - 2)A_{\Delta}(k_{L}, k_{R})/B_{\Delta}(k_{L}, k_{R})$, should satisfy the following recurrence relations:
\begin{subequations}
\begin{align}
G_{\Delta}(k_{L}, k_{R})
&= 	\frac{\frac{\Delta}{2} - 1 \pm ik_{L}}{-\frac{\Delta}{2} \pm ik_{L}}
	G_{\Delta}(k_{L} \pm i, k_{R}), \label{eq:4.3a}\\
G_{\Delta}(k_{L}, k_{R})
&= 	\frac{\frac{\Delta}{2} - 1 \pm ik_{R}}{-\frac{\Delta}{2} \pm ik_{R}}
	G_{\Delta}(k_{L}, k_{R} \pm i), \label{eq:4.3b}
\end{align}
\end{subequations}
These recurrence relations are linear such that they are easily solved by iteration.
But how should we identify the solutions to these recurrence relations with the retarded and advanced two-point functions?
A standard prescription to get the retarded (advanced) two-point functions via AdS/CFT is to use the solution to the Klein-Gordon equation that satisfies the in-falling (out-going) boundary conditions at the horizon $x=\infty$ \cite{Son:2002sd}.
Here we present an alternative approach to get the retarded and advanced two-point functions without knowing the boundary conditions at the horizon $x=\infty$.
A key is the generic causal properties of two-point functions: The retarded two-point function has support only on the future light-cone, whereas the advanced two-point function has support only on the past light-cone.
Let us first focus on the case where the point $(\tau, \theta)$ on the AdS$_{3}$ boundary ($\partial\text{AdS}_{3}$) lies inside the future light-cone $x_{L} = \tau + \theta > 0$ and $x_{R} = \tau - \theta > 0$.
In this case the state $(A_{-}^{0})^{n}(B_{-}^{0})^{m}\phi_{\Delta, k_{L}, k_{R}}^{0} \propto \mathrm{e}^{-i(k_{L} - in)x_{L}}\mathrm{e}^{-i(k_{R} - im)x_{R}}$ converges as $n,m \to \infty$ such that $(A_{-}^{0})^{n}(B_{-}^{0})^{m}\phi_{\Delta, k_{L}, k_{R}}^{0}$ would be well-defined.
Hence it would be natural to expect that the ladder equations $A_{-}^{0}\phi_{\Delta, k_{L}, k_{R}}^{0} \propto \phi_{\Delta, k_{L} - i, k_{R}}^{0}$ and $B_{-}^{0}\phi_{\Delta, k_{L}, k_{R}}^{0} \propto \phi_{\Delta, k_{L}, k_{R} - i}^{0}$ would lead to the retarded two-point function.
Indeed, iterative use of the relations $G_{\Delta}(k_{L}, k_{R}) = \frac{\frac{\Delta}{2} - 1 - ik_{L}}{-\frac{\Delta}{2} - ik_{L}}G_{\Delta}(k_{L} - i, k_{R})$ and $G_{\Delta}(k_{L}, k_{R}) = \frac{\frac{\Delta}{2} - 1 - ik_{R}}{-\frac{\Delta}{2} - ik_{R}}G_{\Delta}(k_{L}, k_{R} - i)$ gives
\begin{align}
G^{R}_{\Delta}(k_{L}, k_{R})
&= 	\frac{\Gamma(\frac{\Delta}{2} - ik_{L})}{\Gamma(1-\frac{\Delta}{2} - ik_{L})}
	\frac{\Gamma(\frac{\Delta}{2} - ik_{R})}{\Gamma(1-\frac{\Delta}{2} - ik_{R})}
	g^{R}(\Delta), \label{eq:4.4}
\end{align}
where $g^{R}(\Delta)$ is a normalization factor given by $g^{R}(\Delta) = \lim_{n,m\to\infty}G^{R}_{\Delta}(k_{L} - in, k_{R} - im)$.
This is the retarded two-point function with desired analytic structure: $G^{R}_{\Delta}(k_{L}, k_{R})$ is analytic in the upper-half complex $k_{L}$- and $k_{R}$-planes and has simple poles at $k_{L} = -i2\pi T(\frac{\Delta}{2} + n)$ and $k_{R} = -i2\pi T(\frac{\Delta}{2} + m)$ ($n,m \in \mathbb{Z}_{\geq0}$) on the lower-half complex $k_{L}$- and $k_{R}$-planes, where $T = \frac{1}{2\pi} (= \frac{1}{2\pi R})$ is the Hawking temperature with respect to the time $\tau$.
Let us next derive the retarded two-point function of CFT$_{2}$ dual to the rotating BTZ black hole \eqref{eq:2.4}.
To this end, let $p_{L}$ and $p_{R}$ be momenta conjugate to the BTZ light-cone coordinates $t \pm \phi$.
Since $\tau \pm \theta$ and $t \pm \phi$ are related as $\tau \pm \theta = (r_{+} \mp r_{-})(t \pm \phi)$, we have $k_{L} = \frac{1}{r_{+} - r_{-}}p_{L}$ and $k_{R} = \frac{1}{r_{+} + r_{-}}p_{R}$, from which we get
\begin{align}
G^{R}_{\Delta}(p_{L}, p_{R})
&= 	\frac{\Gamma(h_{L} - \frac{ip_{L}}{2\pi T_{L}})}{\Gamma(\Bar{h}_{L} - \frac{ip_{L}}{2\pi T_{L}})}
	\frac{\Gamma(h_{R} - \frac{ip_{R}}{2\pi T_{R}})}{\Gamma(\Bar{h}_{R} - \frac{ip_{R}}{2\pi T_{R}})}
	g^{R}(\Delta), \label{eq:4.5}
\end{align}
where $T_{L}$ and $T_{R}$ are the Hawking temperature for left- and right-moving sectors with respect to the BTZ time $t$ and given by
\begin{align}
T_{L}
= 	\frac{r_{+} - r_{-}}{2\pi}
\quad\text{and}\quad
T_{R}
= 	\frac{r_{+} + r_{-}}{2\pi}. \label{eq:4.6}
\end{align}
$h_{L}$ and $h_{R}$ are conformal weights for a scalar operator of dual CFT$_{2}$ given by
\begin{align}
h_{L} = h_{R} = \frac{\Delta}{2}
\quad\text{with}\quad
\Bar{h}_{L} = \Bar{h}_{R} = 1 - \frac{\Delta}{2}. \label{eq:4.7}
\end{align}
Notice that Eq.~\eqref{eq:4.5} precisely coincides with the known results \cite{Birmingham:2001pj} (see also \cite{Iqbal:2009fd,Balasubramanian:2010sc} for the case of fermionic operators.)

Let us next move on to the case where the point $(\tau, \theta) \in \partial\text{AdS}_{3}$ lies inside the past light-cone $x_{L} = \tau + \theta < 0$ and $x_{R} = \tau - \theta < 0$.
In this case the state $(A_{+}^{0})^{n}(B_{+}^{0})^{m}\phi_{\Delta, k_{L}, k_{R}}^{0} \propto \mathrm{e}^{-i(k_{L} + in)x_{L}}\mathrm{e}^{-i(k_{R} + im)x_{R}}$ converges as $n,m \to \infty$ such that $(A_{+}^{0})^{n}(B_{+}^{0})^{m}\phi_{\Delta, k_{L}, k_{R}}^{0}$ would be well-defined.
Iterative use of the relations $G_{\Delta}(k_{L}, k_{R}) = \frac{\frac{\Delta}{2} - 1 + ik_{L}}{-\frac{\Delta}{2} + ik_{L}}G_{\Delta}(k_{L} + i, k_{R})$ and $G_{\Delta}(k_{L}, k_{R}) = \frac{\frac{\Delta}{2} - 1 + ik_{R}}{-\frac{\Delta}{2} + ik_{R}}G_{\Delta}(k_{L}, k_{R} + i)$ then gives the advanced two-point function
\begin{align}
G^{A}_{\Delta}(p_{L}, p_{R})
&= 	\frac{\Gamma(h_{L} + \frac{ip_{L}}{2\pi T_{L}})}{\Gamma(\Bar{h}_{L} + \frac{ip_{L}}{2\pi T_{L}})}
	\frac{\Gamma(h_{R} + \frac{ip_{R}}{2\pi T_{R}})}{\Gamma(\Bar{h}_{R} + \frac{ip_{R}}{2\pi T_{R}})}
	g^{A}(\Delta), \label{eq:4.8}
\end{align}
where $g^{A}(\Delta) = \lim_{n,m\to\infty}G^{A}_{\Delta}(p_{L} + in, p_{R} + im)$.
We note that, since in general the retarded and advanced two-point functions are related by complex conjugate $G_{\Delta}^{A}(p_{L}, p_{R}) = [G_{\Delta}^{R}(p_{L}, p_{R})]^{\ast}$, the normalization constants must be related by $g^{A}(\Delta) = [g^{R}(\Delta)]^{\ast}$.

\subsection*{Acknowledgement} \label{sec:acknowledgement}
The author is supported in part by ESF grant CZ.1.07/2.3.00/30.0034.

\bibliographystyle{utphys}
\bibliography{Bibliography}

\end{document}